\documentclass[runningheads,a4paper]{llncs}

\usepackage{amssymb}
\usepackage{graphicx}
\usepackage{nth}
\usepackage{cite}
\usepackage{wasysym}

\usepackage{fancyvrb}

\usepackage{url}
\usepackage{epstopdf}
\urldef{\mailsa}\path|stys@jcu.cz|
\newcommand{\keywords}[1]{\par\addvspace\baselineskip
\noindent\keywordname\enspace\ignorespaces#1}

\begin{document}

\mainmatter
\title{Model of the Belousov-Zhabotinsky reaction -- Part 2: Central role of the internal excitation growth rate}
\titlerunning{Spiral and diffusion-like behavior in the B-Z reaction}
\author{Dalibor \v{S}tys \and Renata Rycht\'{a}rikov\'{a} \and Tom\'{a}\v{s} N\'{a}hl\'{i}k \and Anna Zhyrova \and \v{S}t\v{e}p\'{a}n Pap\'{a}\v{c}ek}

\authorrunning{Dalibor \v{S}tys et al.}

\institute{Institute of Complex Systems, Faculty of Fisheries and Protection of Waters,\\
University of South Bohemia, Z\'{a}mek 136, 373 33 Nov\'{e} Hrady, Czech Republic
\mailsa\\
\url{http://www.frov.jcu.cz/cs/ustav-komplexnich-systemu-uks}}

\toctitle{Lecture Notes in Bioinformatics}
\tocauthor{Dalibor \v{S}tys et al.}
\maketitle

\begin{abstract}

The model of the Belousov-Zhabotinsky reaction, so-called hodgepodge machine, is discussed in detail and compared with the experimentally determined system trajectory.
We show that many of the features observed in the experiment may be found at different internal excitation growth rates, the $g/maxState$ ratios.
The $g/maxState$ ratio determines the structure of the limit set and defines also details of the system state-space trajectory. While the limit set identical to the experiment and the model may be found, there is not a single $g/maxState$ level which defines the trajectory identical to the experiment. We also propose that there is an inherent experimental time unit defined by the time extent of the bottleneck chemical reaction process which defines the $g/maxState$ ratio and the spatial element of the process.

\end{abstract}

\section{Introduction}

Until recently, the reaction-diffusion simulations based on PDEs,
which are central models of chemical processes, have been used
extensively to explain the self-organizing Belousov-Zhabotinsky
(B-Z) reaction \cite{VanagandEpstein2009}. This kind of simulation
expects an instantaneous chemical change. However, in reality, any
chemical reaction is a combination of the physical breaking of
chemical bonds, splitting, and diffusion of new chemical moieties
over a time span -- a defined elementary timestep needed for the
progression of the spatial-limited chemical process. Therefore, for
the description of the B-Z reaction, we chose a kind of cellular
automaton -- a hodgepodge machine, see e.g. \cite{Gerhardt1989} --
for the time-spatially discrete simulation.

In brief, we consider four processes in the hodgepodge machine algorithm which describe the behavior of the chemical reaction at the quantum (electron) scale~\cite{Stysetal2015}:
\begin{enumerate}
\item Deexcitation -- phase transition -- bond breakage upon reaching the maximal level,
\item spreading the "infection" from neighboring cells,
\item excitation process which is simulated by the acquisition of cell levels (energy transfer) from the average of neighboring cells
and,
\item growth of the excitation inside the cell.
\end{enumerate}
Items 1--2 are the most speculative processes, but we have at least
the experimental evidence for them. Further, we assume that
processes 3 and 4 occur on the border and inside of the structural
element, respectively. In this aspect, the hodgepodge machine may be
considered as the most elementary example with (a) only forward
(excitation) processes and (b) a linear increase of the process
inside the spatial element and on its border. The automaton depends
upon 4 parameters, $maxState$, $k_1$, $k_2$, $g$,  as well as on
initial (an ignition phase) and boundary conditions (usually
periodic), cf. \cite{Stysetal2015}.

In our previous paper \cite{Stysetal2015}, we described the rule for the ignition phase of the hodgepodge machine which led to the course and final state highly similar to that observed in the B-Z reaction \cite{Belousov1959, Zhabotinsky1964, Zhabotinsky1964b}:
\medskip \newline "Let $k_1$ and $k_2$ be a weight of the cell in the ignition state from the interval (0, $maxState$) and that in the ignitial state $maxState$, respectively. Then, if $1/k_{1} + 0/k_{2} < 0.5$ or $0/k_{1} + 1/k_{2} < 0.5$, a limit set of waves and spirals is observed. In all other cases we observe filaments." \medskip

In this article, we continue our research on the hodgepodge machine,
studying the influences of the parameter $maxState$ setting the
maximal state value,  and the internal excitation growth rate $g$,
i.e. the number of levels increased in one simulation step. More
precisely, we aim to understand how the ratio $g/maxState$
influences  the processes under the items 3--4.

\section{Methods}

We used the Wilensky NetLogo hodgepodge model \cite{Wilensky2003}, which was modified in two ways to improve its similarity to the B-Z experiment \cite{Stysetal2015, url1}. The first change (the generation of the random-exponential noise) was introduced to ensure that the simulation started with a small number of ignition points. The second modification (the rounding of the weighted occupied neighboring cells) was implemented in order to allow us to start from these few points, where the original rule (the calculation in integers) did not. The simulation was performed on a canvas of 1000 $\times$ 1000 cells to avoid non-idealities caused by the periodic boundary conditions. Here, we present behavior of the simulation for $k_1=k_2=3$.

For comparison to the simulation, the B-Z reaction was performed as previously published \cite{ZhyrovaStys2014} with the only modification in that we shook (1400 rpm) a big Petri dish ($\diameter$ 200 mm).

In the discussion, we used Wuensche's terminology \cite{Wuensche2011}. The final states of the B-Z reaction and hodgepodge simulation are considered to be limit sets, and the courses of the experiment and its simulation are called trajectories through the state space.

\section{Results and Discussion}
\subsection{Influence of the maximal number of cell states $maxState$ on the limit sets}

\begin{figure}
\centering
\includegraphics[width=0.6\textwidth]{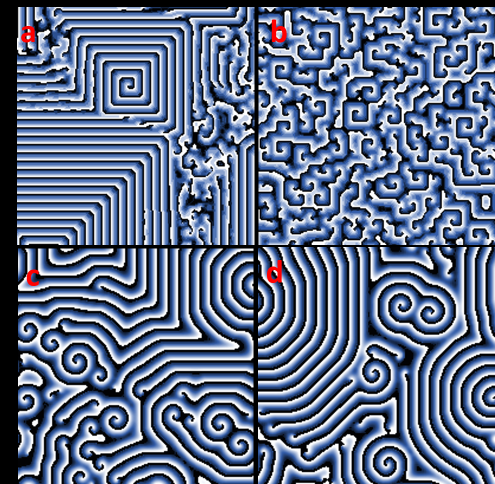}
\caption{The influence of the number of cell states $maxState$ on the limit sets at similar $g/maxState$ ratios of 1:7 (\textbf{a}), 3:20 (\textbf{b}), 28:200 (\textbf{c}), and 280:2000 (\textbf{d}). Total overviews of the limit sets' structures.}
\label{fig:Number_of_levels}
\end{figure}

Multilevel cellular automata are often analyzed for at most 8 cell states, e.g. \cite{Wuensche2011}. Here, we tested the modified Wilensky hodgepodge machine algorithm by changing the maximal number of the achievable cell states $maxStates$ (Fig. \ref{fig:Number_of_levels}).

The limit set -- a mixture of waves and spirals -- highly similar to those observed in the B-Z experiment was achieved at the $g/maxState$ ratio of 28:200 (compare Figs. \ref{fig:Number_of_levels}\textbf{c} and \ref{fig:Number_of_levels_cut}\textbf{c} with Fig. \ref{fig:Experimental_trajectory} in 210 s). We observed branched spirals there (Fig. \ref{fig:Number_of_levels}\textbf{c}). At the same ratio and higher $maxState$ value (Figs. \ref{fig:Number_of_levels}\textbf{d}), the spirals were even more smooth. To obtain these results, we kept the ratio constant and simplified it to 7:50. The simulation at the $g/maxState$ ratio of 3:20 brought up the limit set, which was also similar to the chemical experiment. The $g/maxState$ ratio of 1:7 (Fig. \ref{fig:Number_of_levels}\textbf{a}) led to the formation of square spirals reported earlier \cite{Wuensche2011, Fischetal1991}. The next ratio 28:200 (close to 1:7) gave results most similar to the experiment. The $g/maxState$ ratio of 1:8 (Fig. \ref{fig:Number_of_levels}\textbf{b}) already showed spirals and waves like the B-Z experiment. For all comonnly studied "small-number" ratios, the simulation and reaction in the limit sets differ. At $maxState$ $<$ 20 the spiral evolutions were already qualitatively similar.

Among the low-number $maxState$ levels, the simulation at the 1:7 ratio was extraordinary due to slow evolution to the limit set. Reaching the limit set took much higher number of simulation steps (37,000) than in other cases (e.g., 2,500 steps at $g/maxState$ = 1:8). Indeed, 8 levels, which correspond to the number of pixels in the Moore neighborhood, already tend to make octagons (read below). Just one number below, 7 levels, could change the system's behavior. However, further research needs to be done to understand this phenomenon.

\subsection{Temporarily organized structures in the initial phases and the limit sets at different $g/maxState$ ratios}

\begin{figure}
\centering
\includegraphics[width=0.7\textwidth]{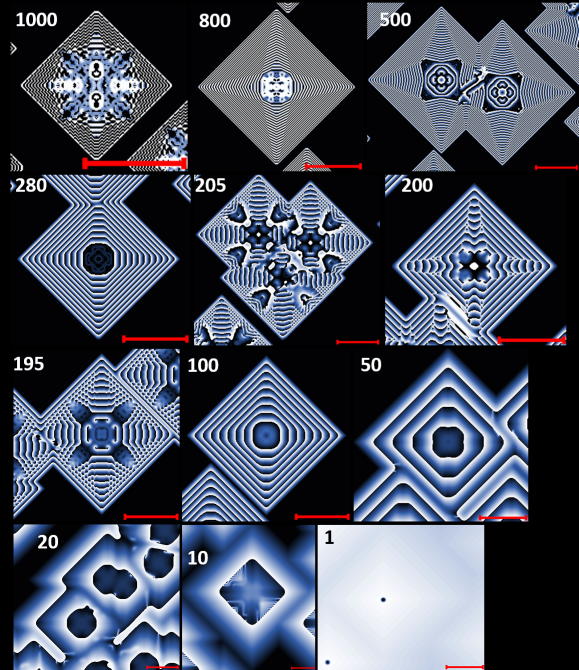}
\caption{The initial states for selected values of $g$ at the
constant $maxState$ = 2000. Each of the scale bars corresponds to
100 px.
 As in the previous figures, the black color corresponds to the state level 0, white color to the maximal state value (set by the constant $maxState$).} \label{fig:typology_of_central_structures}
\end{figure}

\begin{figure}
\centering
\includegraphics[width=0.85\textwidth]{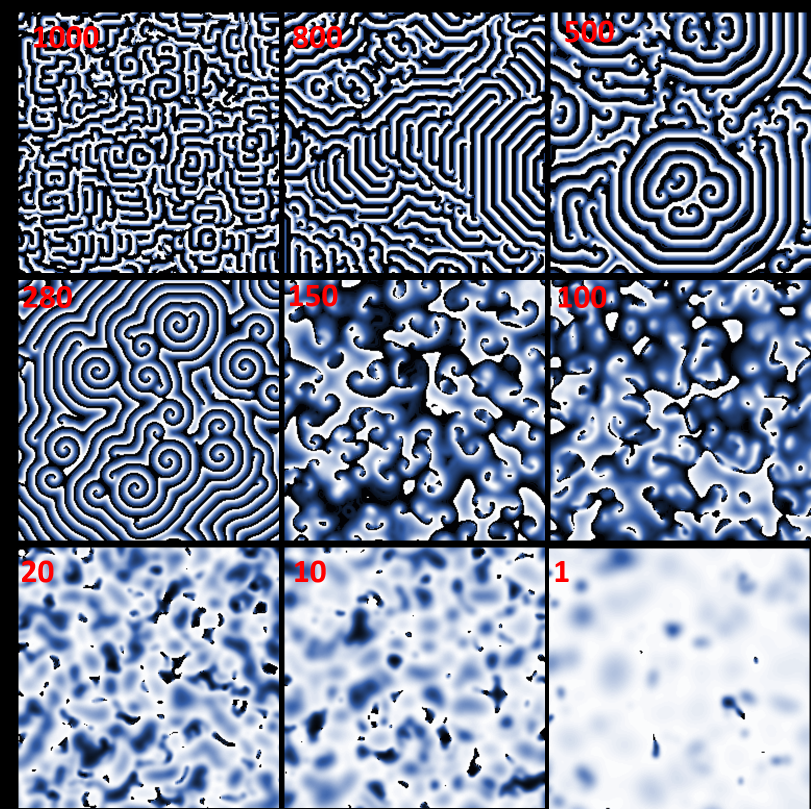}
\caption{The limit sets for selected $g$ at $maxState$ = 2000. From
the upper left to lower right image, the values of $g$ are  1000,
800, 500, 280, 150, 100, 20, 10, and 1, respectively.}
\label{fig:typology_of_limit_sets}
\end{figure}

The temporarily organized structures on the state-space trajectory of the simulation are those on one of the paths of the discrete dynamic network \cite{Wuensche2011}. In Figs. \ref{fig:typology_of_central_structures} and \ref{fig:typology_of_limit_sets} we show the shape dependence of the temporarily organized structures in the initial phases and limit sets on the $g$ value at the constant $maxState$ value (i.e., on the decreasing $g/maxState$ ratio).

Various ignition points, which are assumed to be Gardens of Eden, gave octagonal structures (Fig. \ref{fig:typology_of_central_structures}), whose interiors were typical for the given Gardens of Eden. At the high $g/maxState$ ratio, after the passage of early waves when the interior octagon became regular, a centrally symmetrical structures appeared in the centre. The shapes of such initial central structures are dependent on the value of the $g/maxState$ ratio. We do not describe here the mechanism how these structures arise.

The early square waves gradually broadened with decreasing\linebreak$g/maxState$ ratio (Fig. \ref{fig:typology_of_central_structures}). The lower the $g/maxState$, the more diffuse the structures appeared. At $g/maxState \leq$ 50:2000 we obtained a diffusive central object surrounded by the spreading wavefronts. With $g/maxState$ of 10:2000, the fuzzy diffusive wavefronts were becoming more sparse. At $g/maxState =$ 1:2000 the second wave stopped evolving and the circular dark central structure appeared inside the dense diffusion.

Thus, the $g/maxState$ ratio can explain the thickness of the early waves and determine the number of states achieved in one time interval. This suggests that the wave thickness and density can, to some extent, explain other phenomena such as the built-in local asymmetry of the space \cite{Stysetal2015}, which ignites all additional phenomena and determines the state space trajectory of the process to its limit sets.

The limit sets (Fig. \ref{fig:typology_of_limit_sets}) first emerge from a fully spiral form at $g/maxState$ = 1000:2000. As $g/maxState$ decrease down to 280:2000, the spirals more and more resemble those found in a natural chemical reaction \cite{Stysetal2015}. The waves were broadening and by $g/maxState$ = 150:2000 mature spirals could no longer be observed. Later, the organized structures were broken into a diffusive mixture of darker and lighter stains. At the limit $g/maxState \rightarrow 0$, we are likely to obtained a uniform intensity. It shows the unsuitability of the reaction-diffusion model for the description of the B-Z reaction \cite{VanagandEpstein2009}.

\subsection{Evolution of the state space trajectory}

\begin{figure}
\centering
\includegraphics[width=0.85\textwidth]{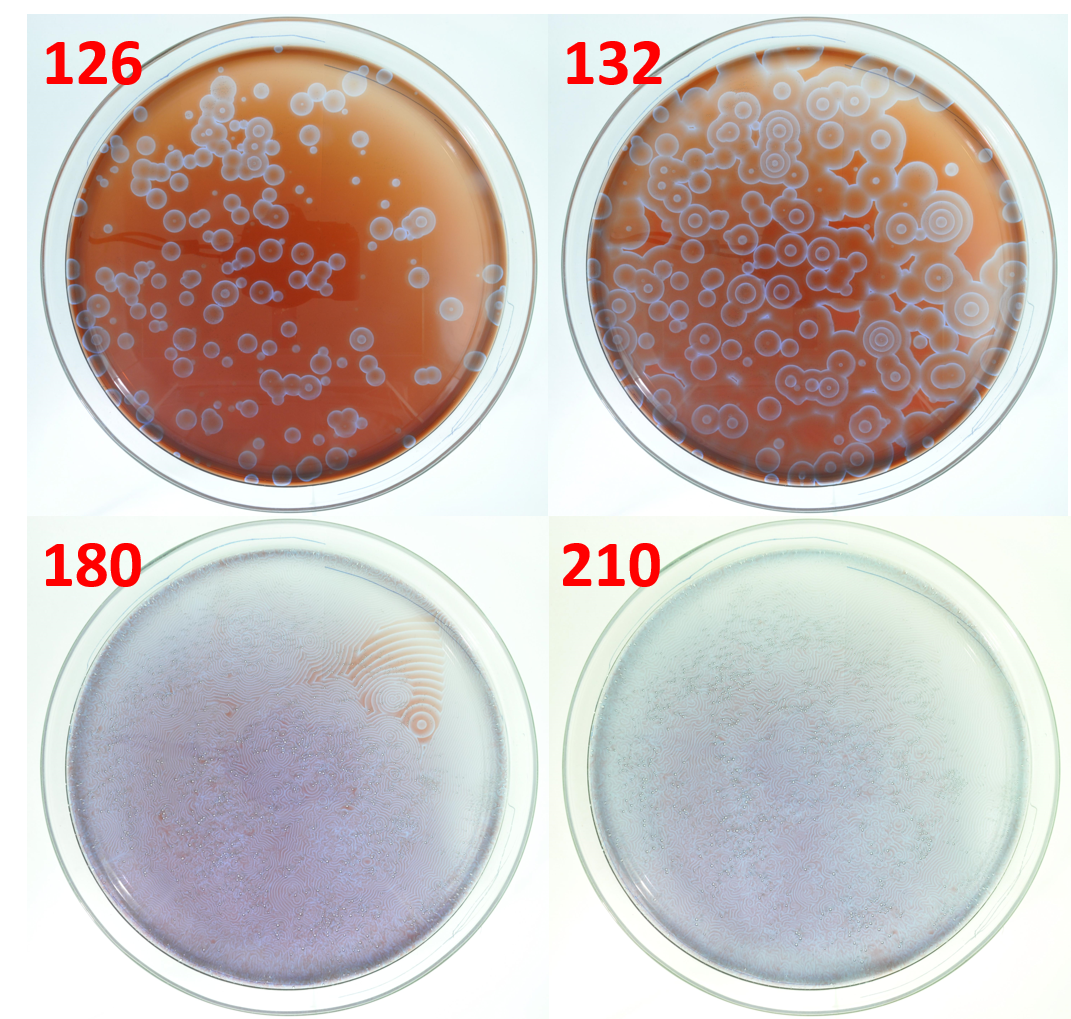}
\caption{Segments of the experimentally determined trajectory of the
Belousov-Zhabotinsky reaction. Numbers correspond to the time
interval (in seconds) from the initial mixing of the reactants.}
\label{fig:Experimental_trajectory}
\end{figure}

The main goal of the hodgepodge machine simulation is to test its consistency with the course of the B-Z reaction (an experimental trajectory). The segments of the reaction are depicted in Fig. \ref{fig:Experimental_trajectory}.

\begin{figure}
\centering
\includegraphics[width=0.85\textwidth]{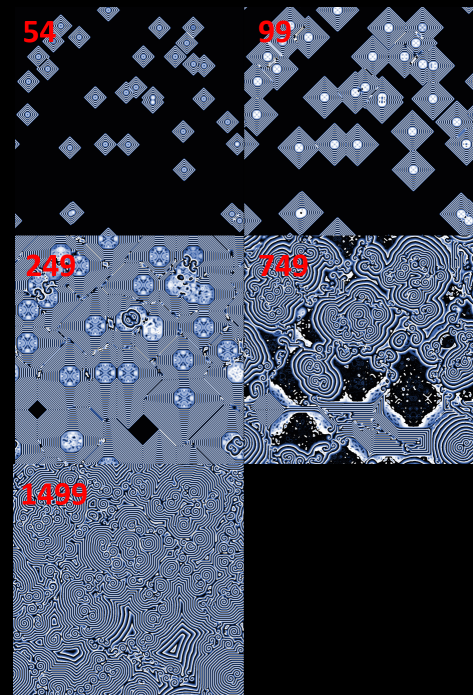}
\caption{The phases of the state-space trajectory at $g/maxState =$
280:2000. \textbf{Step 54} -- The octagonal waves in the centre
collapses and is replaced by the fractal structure. \textbf{Step 99}
-- In the places of the collision of two central structures, early
spiral structures arise. \textbf{Step 249} -- Spiral structures are
being surrounded by the elliptical wavefront. \textbf{Step 749} --
The simulation canvas is being filled by new spiral waves.
\textbf{Step 1499} -- Structures become almost regular by
"compression" of waves of various origins.}
\label{fig:Typical_state_trajectory_phases}
\end{figure}

Despite the similarity of the limit set between the simulation at $g/maxState$ ratio = 280:2000 (Figs. \ref{fig:Number_of_levels}\textbf{c} and \ref{fig:Number_of_levels_cut}\textbf{c}) and the chemical experiment, the trajectories towards these limit sets were a bit different. In the case of the simulation (Fig. \ref{fig:Typical_state_trajectory_phases}), after the ignition, the structures evolved in a sequence of square waves with round centers. After that, the first spiral doublet arose in the vicinity of two central objects such that the waves collided in one point. The doublet soon became surrounded by an elliptical wave thicker than the early square wave. This elliptical wave system had an expanding "diffusive" character until it was compressed by another system of elliptical waves which slowly evolved at other points.

In case of the model, early circular waves analogous to those in the experiment, were achieved only at very low $g/maxState$ ratios, when the waves were broad and the spirals were not formed (e.g., Fig. \ref{fig:typology_of_central_structures} for $g$ = 20). This suggests that in the early phase of the experimental trajectory, there is a second process which brings the evolution of a different trajectory. Later, when this process is exhausted, the evolution of the system follows the path described by the hodgepodge machine model.

\section{Conclusions}

This paper examines the suitability of the hodgepodge machine simulation for the description of the Belousov-Zhabotinsky self-organizing reaction. The reason for doing so is that the hodgepodge machine represents a time-spatial discretization. The time-spatial discretization is the idealized process, where the time element determines (a) a set of events achieved at the time within a spatial element (a cell) -- the $g$ constant -- and (b) the rest of events on the spatial element's (cell's) borders -- the $maxState$ constant. We may consider analogies for such behaviour in known energy transfer processes, i.e. the difference between resonance energy transfer and energy transfer in systems of overlapping orbitals.

The analysis of these constants showed that the $g/maxState$ ratio determined the thickness of waves and the course of the state space trajectory irrespective to the total $maxState$ value.
 The decreasing $g/maxState$ ratio led to the loss of the wave structure, which eventually resulted in a fully diffusive picture. There also exists a lower limit of $maxState$ for the appearance successful course of the simulation. The higher the $maxState$ value, the smoother the spirals were.

\subsubsection*{Acknowledgments.}
This work was partly supported by the Ministry of Education, Youth and Sports of the Czech Republic -- projects CENAKVA (No. CZ.1.05/2.1.00/01.0024) and CENAKVA II (No. LO1205 under the NPU I program), by Postdok JU CZ.1.07/2.3.00/30.0006, and GAJU Grant (134/2013/Z 2014 FUUP). Additional thank to Petr Jizba and Jaroslav Hlinka for important discussions and to Kaijia Tian for edits.

\end{document}